\documentclass[groupaddress,twocolumn,showpacs,preprintnumbers,amsmath,amssymb,pra,10pt,aps]{revtex4-1}
\usepackage{graphicx}
\usepackage{dcolumn}
\usepackage{bm}
\usepackage{subfigure}
\usepackage{multirow}
\usepackage{soul}
\usepackage{float}
\usepackage[normalem]{ulem}
\usepackage[usenames,dvipsnames]{color}
\usepackage{bbold}

\definecolor{tangerine}{rgb}{0.944,0.522,0}

\newcommand{\editor}[2]{%
  \expandafter\newcommand\csname #1note\endcsname[1]{%
    \textcolor{#2}{(\textbf{#1:} ##1)}}%
  \expandafter\newcommand\csname #1\endcsname[1]{%
    \textcolor{#2}{##1}}%
  \expandafter\newcommand\csname #1cancel\endcsname[1]{%
    \textcolor{#2}{\sout{##1}}}%
  \expandafter\newcommand\csname #1change\endcsname[2]{%
    \textcolor{#2}{\sout{##1} ##2}}%
  \newenvironment{#1text}{\color{#2}}{\color{black}}
}

\begin{document}

\title{Rationalizing doping and electronic correlations in LaFe$_2$As$_2$}
\author{Tommaso Gorni}
\author{Diego Florez-Ablan}
\author{Luca de'~Medici}
\affiliation{LPEM, ESPCI Paris, PSL Research University, CNRS, Sorbonne Universit\'e, 75005 Paris France}

\date{\today}

\begin{abstract}
We compute the electronic properties of the normal state of uncollapsed LaFe$_2$As$_2$, taking into account local dynamical correlations by means of slave-spin mean-field+density-functional theory.
Assuming the same local interaction strength used to model the whole electron- and hole-doped BaFe$_2$As$_2$ family, our calculations reproduce the experimental Sommerfeld specific heat coefficient, which is twice the value predicted by uncorrelated band theory.
We find that LaFe$_2$As$_2$ has a reduced bare bandwidth and this solves the apparent paradox of its sizeable correlations despite its nominal valence d$^{6.5}$, which would imply extreme overdoping and uncorrelated behaviour in BaFe$_2$As$_2$.
Our results yield a consistent picture of the whole 122 family and point at the importance of the correlation strength, rather than sheer doping, in the interpretation of the phase diagram of iron-based superconductors

\end{abstract}

\pacs{}

\maketitle

The conventional phase diagram of iron-based superconductors (IBSC) typically displays three zones: the paramagnetic metallic parent compound, an antiferromagnetic metal obtained upon lowering the temperature, and a superconducting phase replacing the magnetic order upon (both electron- and hole-) doping or under pressure~\cite{revJohnston:2010,revJohnson:2015,revMartinelli:2016}. 
Excluding the notable exceptions of, e.g., FeSe and LiFeAs (which lack the magnetic phase), such a phase diagram is shared among different IBSC families~\cite{revJohnston:2010,revStewart:2011,revJohnson:2015,revMartinelli:2016,Dai:2015} and it has been thoroughly studied in the 122 family by leveraging on the ease at doping the BaFe$_2$As$_2$ parent compound with holes in the Ba$_{1-x}$K$_x$Fe$_2$As$_2$ series, and with electrons in the Ba(Fe$_{1-x}$Co$_x$)$_2$As$_2$ series~\cite{Hardy:2010,Hardy:2016}.
These studies gave evidence of a strong asymmetry between the hole- and electron-doped side: in the former the superconducting dome extends up to $0.5$--$0.6$ extra holes per unit cell, while in the latter the superconducting critical temperature ($T_c$) drops down to zero before reaching $0.15$ extra electrons per Fe atom~\cite{revJohnson:2015,revMartinelli:2016}.

Recent studies on the (uncollapsed) La$_{0.5-x}$Na$_{0.5+x}$Fe$_2$As$_2$ series, where hole- and electron-doping are attained by acting only on the intercalated layer, initially showed similar results on the hole-doped ($x>0$) side, where a superconducting phase emerges from an antiferromagnetic state around $x\approx 0.15$ and reaches the maximum $T_c = 27.0$\,k at $x\approx 0.3$~\cite{Iyo:2018}. 
%	
%The drop of $T_c$ for $x > 0.35$ could not be measured due to the impossibility to synthetize samples in that regime.
%
The later discovery of superconductivity in the electron-doped end member of the series, LaFe$_2$As$_2$, with $T_c\approx 12.1$\,k, came therefore as a surprise, the compound corresponding to a nominal valence of $6.5$ electrons for the Fe ion~\cite{Iyo:2019}, that is a doping of 0.5 electrons from the valence of the reference parent compounds in all iron-based superconductors, like BaFe$_2$As$_2$. There, at this extreme electron overdoping superconductivity has long disappeared. 

Such a finding has been initially explained via band-structure analysis by means of DFT(GGA) calculations, showing that the strong hybridization between the La-$5d$ and the Fe-$3d$ levels conceals a hole pocket generated by Fe bands, suggesting a smaller electronic doping, estimated in the range of $0.25$--$0.40 e^-$/Fe~\cite{Usui:2019,Mazin:2019}.  
%
%This analysis leverages on the hypothesis that spin fluctuations drive the superconducting transition and that La-$5d$ do not take part to those, notwithstanding their strong hybridization with the Fe-$3d$ orbitals.
%
However, band structures derived from LDA and GGA functionals are known to yield poor a description of electronic quasiparticles in IBSC, mainly due to the underestimation of local dynamical correlations, leading to at least a factor-two discrepancy between the computed and experimental electronic bandwidths~\cite{Coldea:2008,Brouet:2009,Lee:2012,Terashima:2013,Watson:2015}.
Even though these effects are supposed to be less relevant on the electron-doped side, nuclear magnetic and quadrupolar resonance (NMR and NQR) studies found that antiferromagnetic spin fluctuations are not completely suppressed in LaFe$_2$As$_2$ (on contrary to what happens in Ba(Fe$_{0.5}$Co$_{0.5}$)$_2$As$_2$)~\cite{Kouchi:2019}, and specific heat measurements have shown a Sommerfeld coefficient roughly a factor 2 bigger than what predicted by DFT(GGA,LDA) calculations~\cite{Pallecchi:2020}.
While the first finding alone leaves the door open to an uncorrelated, smaller effective doping scenario, when considered together with the specific heat measurement it points more likely to a decisive role played by electronic correlations in LaFe$_2$As$_2$.
More accurate treatments of these dynamical correlation effects, by means, e.g., of Dynamical Mean-Field Theory (DMFT) or slave-particle approaches, have indeed been shown to generally provide electronic bandwidths in much closer agreement with experimental findings in IBSC, and generally improve the description of their low-energy electronic properties~\cite{Yin:2011,deMedici:2014,Gorni:2021}.
DMFT studies have been performed on LaFe$_2$As$_2$ and CaFe$_2$As$_2$ compounds in order to inspect the structural and doping dependence of the superconducting transition, coming however to rather different conclusions, most likely due to the usage of different structural parameters and interaction-strength values~\cite{Acharya:2020,Zhao:2021}.

Here, we study LaFe$_2$As$_2$ electronic properties by deriving a realistic tight-binding from DFT(GGA) calculations, and taking into account local dynamical correlations by  means of slave-spin mean-field (SSMF) theory.
The numerical agility of our approach allows us to explore the interaction-strength space and pinpoint the interaction value by comparing against recent specific heat measurements~\cite{Pallecchi:2020}.
We find that the experimental Sommerfeld coefficient of LaFe$_2$As$_2$ is reproduced by using the same interaction values which captures its evolution throughout a large variety of electron- and hole-doped BaFe$_2$As$_2$~\cite{Hardy:2016}, so placing LaFe$_2$As$_2$ in the same theoretical framework of other 122 compounds.
We compare the series of stochiometric compounds XFe$_2$As$_2$ with X=K,Ba,La and indeed correlations decrease drastically with increasing filling (5.5, 6.0 and 6.5 electrons/Fe respectively), as expected in the scenario positing the influence of a Mott insulator that would be realized at half-filling (i.e. 5.0 electrons/Fe) and extending its influence over a large range of electronic densities\cite{deMedici:2014}.
However further inspection of the model reveals that correlations are stronger in LaFe$_2$As$_2$ with respect to a simulation of BaFe$_2$As$_2$ doped with electrons so to reach the same nominal valence of 6.5 electrons, mainly due to the smaller bare bandwidth of the former.
We argue therefore that the apparent paradox of superconductivity at high electron doping in LaFe$_2$As$_2$ can be solved by analysing the phase diagram in terms of the strength of local correlations, rather than sheer doping. 

\section{DFT calculations}
%
%%%%%%%%%%%%%%

The DFT calculations with the PBE functional have been carried out with the {\sc Quantum ESPRESSO} package~\cite{Giannozzi:2009,Giannozzi:2017}, using norm-conserving pseudopotentials from the PseudoDojo library~\cite{vanSetten:2018} and a plane-wave cutoff of $\epsilon_{\rm cut} = 90$\,Ry. Integrals over the Brillouin zone have been converged with a gaussian smearing of $\sigma = 0.01$\,Ry and a $\Gamma$-centred uniform grid of $8\times 8 \times 8$ points.
The crystal structures have been fixed to the experimental values, resumed in Tab.~\ref{tab:crystal-structure}.

%%%%%%%%%
\begin{table}[h]
%%%%%%%%%
%
\centering
\begin{tabular}{c|c|c|c}
                      &   LaFe$_2$As$_2$~\cite{Iyo:2019}     &     BaFe$_2$As$_2$~\cite{VillarArribi:2018}      &      KFe$_2$As$_2$~\cite{VillarArribi:2018}     \\
 \hline\rule{0mm}{4mm}
$a$ ($\AA$)   &    3.9376                                              &     3.9625                                   &                   3.844                     \\[2pt]
$c$ ($\AA$)   &   11.7317                                             &    13.0168                                  &                  13.916                     \\[2pt]
$z_{\rm As}$  &    0.3657                                             &      0.3545                                  &                    0.35249                 \\  
\end{tabular}
\caption{\label{tab:crystal-structure}
Structural parameters of the 122 family used in our numerical simulations, corresponding to the experimental uncollapsed body-centered tetragonal phase. The $z_{\rm As}$  coordinate is reported in units of the $c$-axis length.
}
%%%%%%%%%
\end{table}
%%%%%%%%%

\begin{figure}[h] 
%%%%%%%%%
%
\centering
\includegraphics[width=\columnwidth,trim=0 0 0 0 clip]{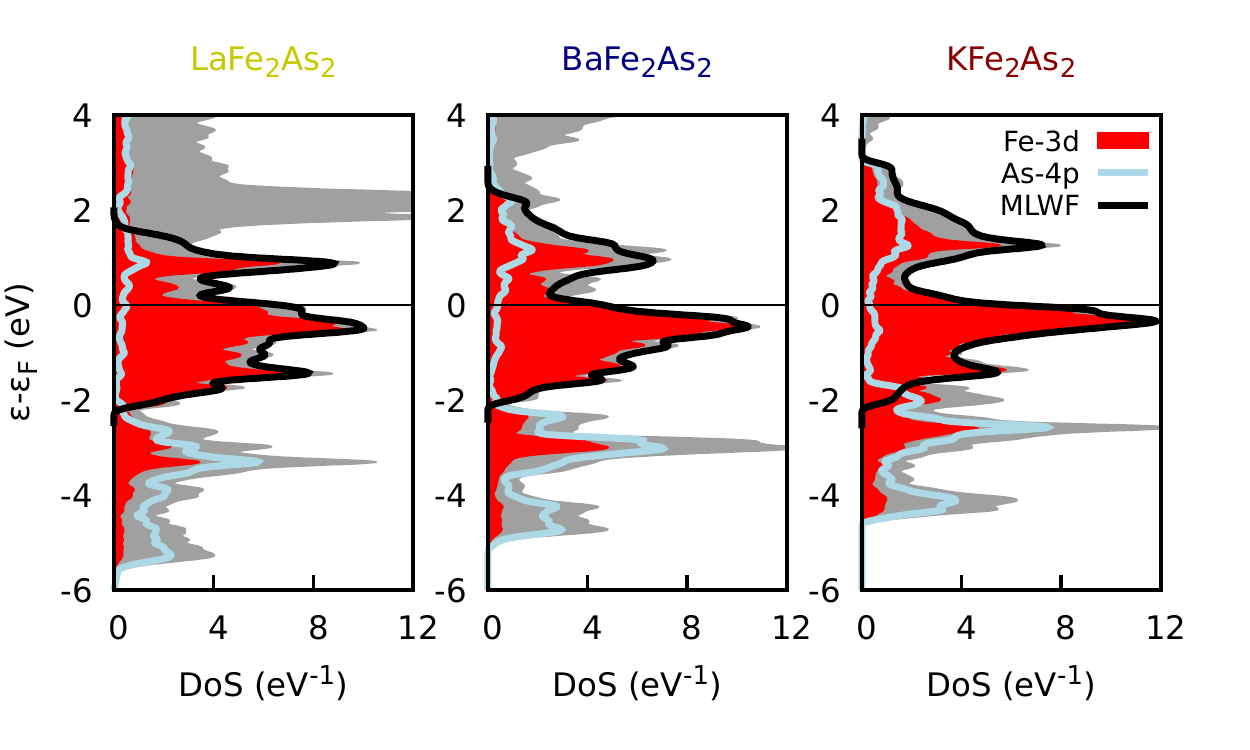}
\caption{\label{fig:pdos}
PBE density of states (DoS) per unit cell of the 122 compounds (grey).
The PBE DoS projected on Fe-3d atomic states is shown in red, while the DoS projected on the As-4p atomic states is shown in light-blue. 
The black line represents the DoS obtained by our maximally-localized Wannier function (MLWF) construction spanning the low-energy region around the Fermi level.
 }
 %%%%%%%
\end{figure}

%
% DFT DoS
%
The PBE density of states (DoS) of LaFe$_2$As$_2$, BaFe$_2$As$_2$ and KFe$_2$As$_2$ are reported in Fig.~\ref{fig:pdos}, together with their projections onto the Fe-$3d$ and As-$4p$ atomic wavefunctions~\footnote{%
As routinely done in planewave+pseudopotential codes, atomic wavefunctions are obtained via a Löwdin orthogonalization of the pseudopotential atomic eigenstates~\cite{SanchezPortal:1995}.
}.
The DoS profiles show the typical structure of IBSC compounds, with low-energy quasiparticles mainly of Fe-$3d$ character mildly hybridized with the subjacent As-$4p$ bands, and are consistent with a hole-doping of the Fe-$3d$ manifold proceeding from LaFe$_2$As$_2$ to KFe$_2$As$_2$, passing through BaFe$_2$As$_2$.
We further notice that the Fe-$3d$ manifold roughly starts around $-2$\,eV in all the three compounds, but has different extensions within the empty-state manifold. While in KFe$_2$As$_2$ the Fe-$3d$ manifold has negligible hybridization with the above empty K-$3d$ bands, it hybridizes  with the Ba-$5d$ empty bands in BaFe$_2$As$_2$, and  with the La-$5d$ bands in LaFe$_2$As$_2$ (conversely the localized La-$4f$ electrons, yielding the narrow peak in the DoS around 2\,eV, have negligible hybridization with the conduction manifold). 
This results into different conduction bandwidths at the PBE level, which can be estimated as $W \approx 5.4$\,eV in KFe$_2$As$_2$, $W \approx 4.7$\,eV in BaFe$_2$As$_2$ and $W \approx 4.1$\,eV in LaFe$_2$As$_2$.

%%%%%%%%%
%%%%%%%

%
% Sommerfeld
%
A relevant quantity that can be extracted from the value of the DoS  at the Fermi level is the Sommerfeld coefficient $\gamma_n = \pi^2 k^2_{\rm B}{\rm DoS}(\varepsilon_{\rm F})/3$, i.e.\ the linear coefficient in the electronic specific heat as a function of temperature, which is a sensitive probe of the renormalized band structure and thus of local dynamical correlations.
As it has been previously shown, the $\gamma_n$ coefficients computed at the PBE level tend to be smaller than the experimental values in the whole 122 family, with mass enhancements ranging from $\gamma^{\rm exp.}_n / \gamma^{\rm PBE}_n \approx 1.2$ for electron dopings of $0.2$, up to $\gamma^{\rm exp.}_n / \gamma^{\rm PBE}_n \approx 10$  for hole dopings of $0.5$ realized i.e. in KFe$_2$As$_2$~\cite{deMedici:2014,Hardy:2016}.
In the case of LaFe$_2$As$_2$, our calculations yield  $ \gamma^{\rm PBE}_n \approx 18$\,mJ\,mol$^{-1}$k$^{-2}$, to be compared against the experimental estimate of $ \gamma^{\rm exp.}_n = (35 \pm 2)$\,mJ\,mol$^{-1}$k$^{-2}$~\cite{Pallecchi:2020}. 
Such a factor-two difference testifies, on the one hand, the relevant impact of local dynamical correlations but, on the other hand, it does not reconcile with the traditional phase diagram of the 122 family, showing vanishing mass enhancements (and no superconducting transition) for electron doping beyond $\approx 0.2$~\cite{Hardy:2010}.

% Exp Sommerfeld
% KFe2As2 -->   91-94      (used 102 in figure, digitized from Luca's plot, from Hardy:2016)                   ~\cite{deMedici:2014,Kim:2011,AbdelHafiez:2012}
% BaFe2As2 --> 27 ?   (used 30 in figure, digitized from Luca's plot)                    27 from extrapolation~\cite{Hardy:2010}
% LaFe2As2 --> 35 +\- 2 mJ mol^{-1} K^{-2}   ~\cite{Pallecchi:2020}

% PBE Sommerfeld
%                       wannier                                  DFT (nscf 16x16x16 sigma=0.01 Ry)
%  KFe2As2 -->  11-17 mol^{-1} K^{-2}                       13.48
% BaFe2As2 --> 11 mJ mol^{-1} K^{-2}                      10.59
% LaFe2As2 --> 17-18 mJ mol^{-1} K^{-2}                 16.67

%
% Wannierization	
%
%%%%%%%%%%%%%%
%
\section{Wannier d-model}
%
%%%%%%%%%%%%%%
In order to address the effect of local dynamical correlations, we project the PBE Bloch states spanning the low-energy manifold on a set of five maximally-localized Wannier functions centered on each Fe site, using the {\sc wannier90} code~\cite{Pizzi:2020}.
Such a procedure is well-defined when applied to a subset of isolated energy bands, however difficulties may arise when the target Bloch states overlap with other energy bands. In these cases, a disentanglement procedure needs to be applied before proceeding with the spread minimization, i.e.\ a subset of energy bands has to be chosen for each $\bm{k}$ point~\cite{Marzari:2012}. 
The typical disentanglement procedure starts from the projection of the Bloch states lying within a larger energy window over a set of trial localized orbitals with a given atomic symmetry, and by orthonormalizing the the resulting set via the Löwdin procedure. The resulting  wavefunctions can be further refined by imposing smoothness in the $\bm{k}$ space~\cite{Souza:2001}. In practice, such a procedure is performed only over a subset of the Bloch states, since the one residing in a smaller energy window are kept frozen, in order to enforce the final local basis set to faithfully reproduce the original bands, at the price of less-localized Wannier functions.
%%
% 
% Ba funzia perché escludo zona ibridizzazione da zona frozen, e la banda fa quel che le pare ma in Lantanio non si può.
 
%
% Spiega meglio come funziona wannierizzazione (spread può perdere simmetria della proiezione iniziale...)
%
In the cases considered here, KFe$_2$As$_2$ presents no particular difficulties and the final set of maximally-localized Wannier functions describes faithfully the conduction manifold.
For what concerns BaFe$_2$As$_2$, the hybridization with the Ba-$5d$ manifold hampers a proper description of empty bands, nonetheless a well-behaved set of Wannier functions is reached with the spread minimization~\cite{Miyake:2010,Graser:2010}.
Finally, in the case of LaFe$_2$As$_2$ these difficulties become severe due to the stronger hybridization with the La-$5d$ bands, and only Wannier projections models including these states explicitly have been considered in the literature so far~\cite{Usui:2019,Mazin:2019,Zhao:2021}.

%%%%%%%%%
\begin{figure}[h] 
%%%%%%%%%
%
\centering
\includegraphics[width=\columnwidth,trim=0 0 15 0 clip]{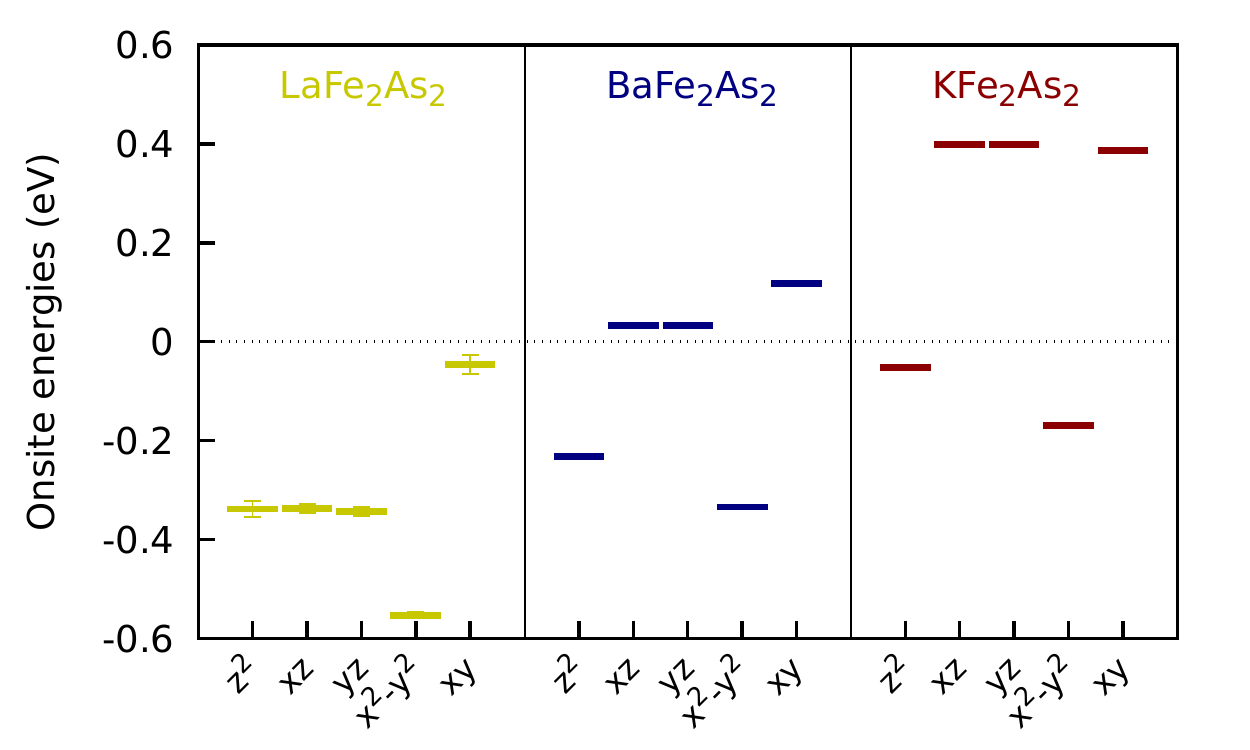}
\caption{\label{fig:onsite-energies}
Onsite energies of our maximally-localized-Wannier-function d-models, referred to the respective Fermi energies.
The error bars represent the discrepancy between the two Fe sites, noticeable only in the case of LaFe$_2$As$_2$.
}
%%%%%%%%%
\end{figure}  
%%%%%%%%%

In the spirit of finding a minimal d-model for the conduction bands, we have pursued this path for LaFe$_2$As$_2$ by carefully inspecting the quality of the tight-binding Hamiltonian as different disentanglement energy windows were considered. 
We noticed that, even though the Bloch states are reproduced quite satisfactorily by different d-models, the disentanglement procedure breaks the equivalence between the two iron sites. Among these, we have therefore chosen the d-model with the minimum difference among the onsite energies of the two equivalent iron atoms, quantified on average  as $\approx 0.02$\,eV and with a maximum deviation of  $\approx 0.04$\,eV in the case of the $xy$ orbital.
The resulting onsite levels are shown in Fig.~\ref{fig:onsite-energies} compared with the ones of BaFe$_2$As$_2$ and KFe$_2$As$_2$, showing an overall consistency within the 122 family and confirming the electron-doped character of the d-manifold in LaFe$_2$As$_2$.
Finally, in Fig.~\ref{fig:pdos} we compare the DoS of our tight-binding models (black lines) with the PBE ones, which validate out preliminary estimates of the d-manifold bandwidths in these compounds.

%
%  KFe2As2 --> Wannier centres precise within 10^-8 Angstrom
% BaFe2As2 --> Wannier centres precise within 10^-8 Angstrom on the xy plane, 0.02 Angtrom along z
% LaFe2As2 --> Wannier centres precise within ~0.06 Angstrom on xy, until 0.4 Angstrom along z
%
% Say in LaFe2As2 we have local hoppings, renormalized with <O+><O> (check).
%

%%%%%%%%%%%%%%
%
\section{Local dynamical correlations}
%
%%%%%%%%%%%%%%

The projection of the PBE Bloch states onto maximally-localized Wannier functions, as discussed in the previous section, yields a tight-binding Hamiltonian in the form of
\begin{equation}
\hat{\mathcal{H}}_0
=
\sum_{\substack{i\neq j \\ m,m',\sigma}}
t^{mm'}_{ij}
\hat{d}^{\dag}_{im\sigma}
\hat{d}_{jm'\sigma}
+
\sum_{i,m,\sigma}
\epsilon_{im\sigma}
\hat{n}_{im\sigma}
\, ,
\label{eq:H0}
\end{equation}
where $\hat{d}^{\dag}_{im\sigma}$ creates an electron in the Wannier spin-orbital $m\sigma$ and lattice site $i$, and $\hat{n}_{im\sigma} = \hat{d}^{\dag}_{im\sigma}\hat{d}_{im\sigma}$ is the corresponding number operator. 

A Hubbard-Kanamori Hamiltonian is used to include the local interactions to be treated in a dynamical many-body fashion
\begin{align}
\hat{\mathcal{H}}%-\mu\hat{N} 
=\: &
\hat{\mathcal{H}}_0 %-\mu \sum_{m\sigma} \hat{n}_{m\sigma}
+
U
\sum_{im} \hat{n}_{im\uparrow}\hat{n}_{im\downarrow}
\nonumber\\
&
+
(U-2J)
\sum_{i,m\neq m'} \hat{n}_{im\uparrow}\hat{n}_{im'\downarrow}
\nonumber\\
&
+
(U-3J)
\sum_{i,m < m',\sigma} \hat{n}_{im\sigma}\hat{n}_{im'\sigma}
\, ,
\label{eq:kanamori}
\end{align}
where U is the intra-orbital Coulomb repulsion and J the Hund coupling~\footnote{%
Here the local off-diagonal terms of the Kanamori Hamiltonian (spin-flip and pair-hopping) are customarily dropped.
}. In this framework the double counting energy for these interactions is absorbed into the chemical potential.

Here we solve the many-body Hamiltonian~\eqref{eq:kanamori} within the slave-spin mean-field theory (SSMF)~\cite{deMedici:2005,Hassan_CSSMF,deMedici:2017}, which has proven to be a robust approximation for IBSC by, e.g., successfully capturing their orbital-differentiation signatures~\cite{deMedici:2014}, or predicting the evolution of the Sommerfeld coefficient upon doping  in the 122 family~\cite{Hardy:2016}.
SSFMT describes the Fermi-liquid low-temperature paramagnetic metallic phase of~\eqref{eq:kanamori} as a quasiparticle Hamiltonian
\begin{equation}
\hat{\mathcal{H}}_{\rm QP}
=
\!\!\!
\sum_{\substack{i\neq j\\ mm'\sigma}}
\!\!
\sqrt{Z_m Z_{m'}}
\,
t^{mm'}_{ij}
\hat{f}_{im\sigma}^{\dag}
\hat{f}_{jm'\sigma}
+
\sum_{im\sigma}
(\epsilon_m -\tilde{\lambda}_m)\hat{n}^f_{im\sigma}
\, ,
\label{eq:H-QP}
\end{equation}
where the (orbital-dependent) quasiparticle renormalizations $Z_m$ and on-site-energy shifts $\tilde{\lambda}_m$ are determined solving the self-consistent slave-spin equations for given values of the local interactions $U$ and $J$~\cite{deMedici:2017}.

%
% Sommerfeld coefficient
%
%%%%%%%%%
\begin{figure}
%%%%%%%%%
%
\centering
\includegraphics[width=0.9\columnwidth,trim=0 0 0 0 clip]{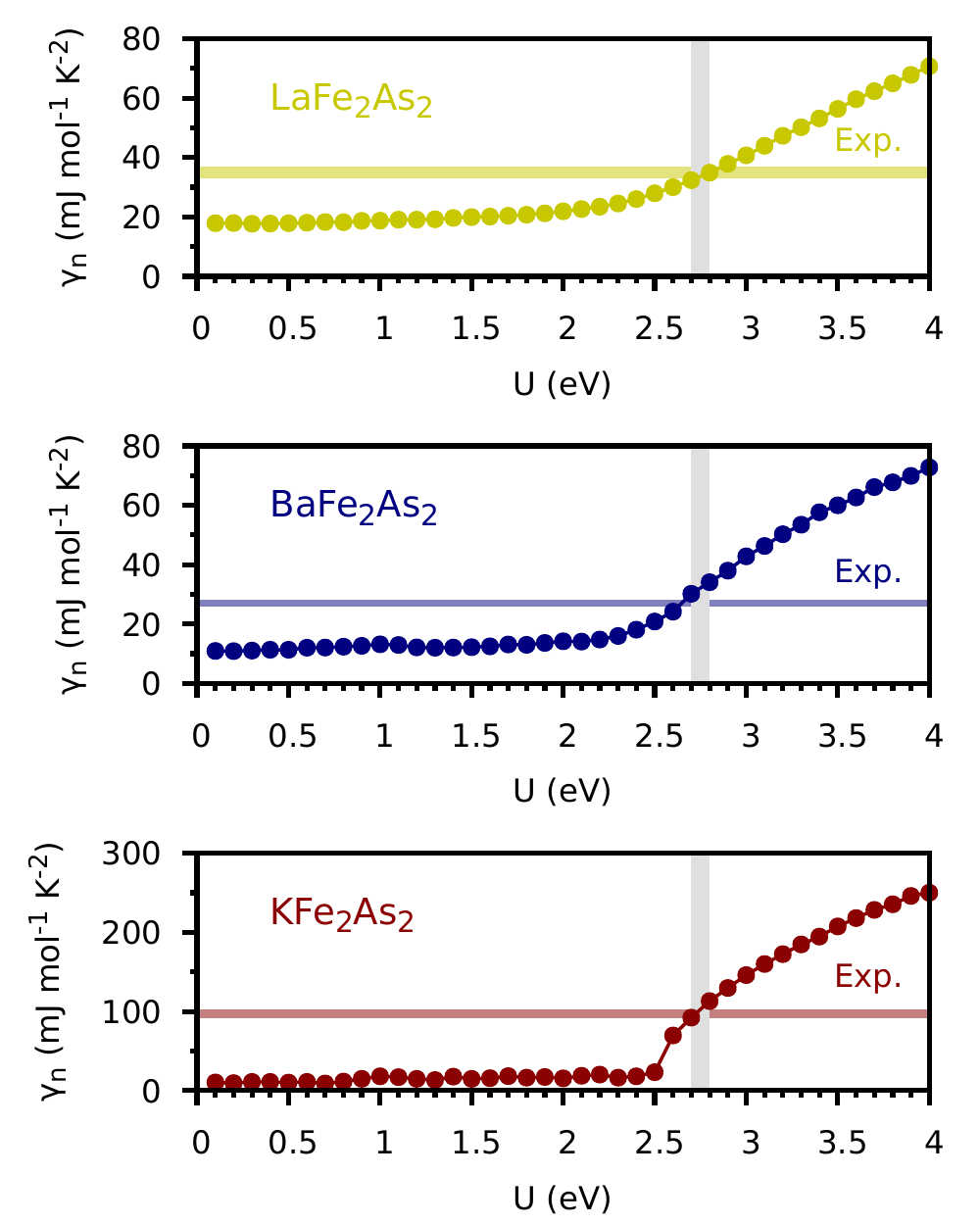}
\caption{\label{fig:sommerfeld}
Sommerfeld coefficients of the 122 family from our SSMF@PBE calculations as a function of the Hubbard $U$ strength ($J/U=0.25$), compared with the experimental values (~\cite{Hardy:2016, Pallecchi:2020}).
The trend over the whole family is captured by $U\approx 2.7\div 2.8$\,eV (grey line). 
The $U=0$ value corresponds to the PBE case.
}
%%%%%%%%
\end{figure}  
%%%%%%%%

In Fig.~\ref{fig:sommerfeld} we report the Sommerfeld coefficient as a function of the intra-orbital Coulomb repulsion $U$, computed from the renormalized DoS in the case of LaFe$_2$As$_2$, BaFe$_2$As$_2$ and KFe$_2$As$_2$. 
In the case of LaFe$_2$As$_2$, we find that the experimental value is reproduced for a local interactions of $U\approx 2.7\div 2.8$\,eV and $J/U =0.25$.
Quite strikingly, these parameters are the same capturing the evolution of the Sommerfeld coefficient along the K$_{x}$Ba$_{1-x}$Fe$_2$As$_2$ series, even accounting for Rb and Cs substitutions~\cite{Hardy:2016}, supporting the robustness of this model for the 122 family.

\begin{figure}[h] 
%%%%%%%%%%
%
\centering
\includegraphics[width=\columnwidth,trim=0 15 0 0 clip]{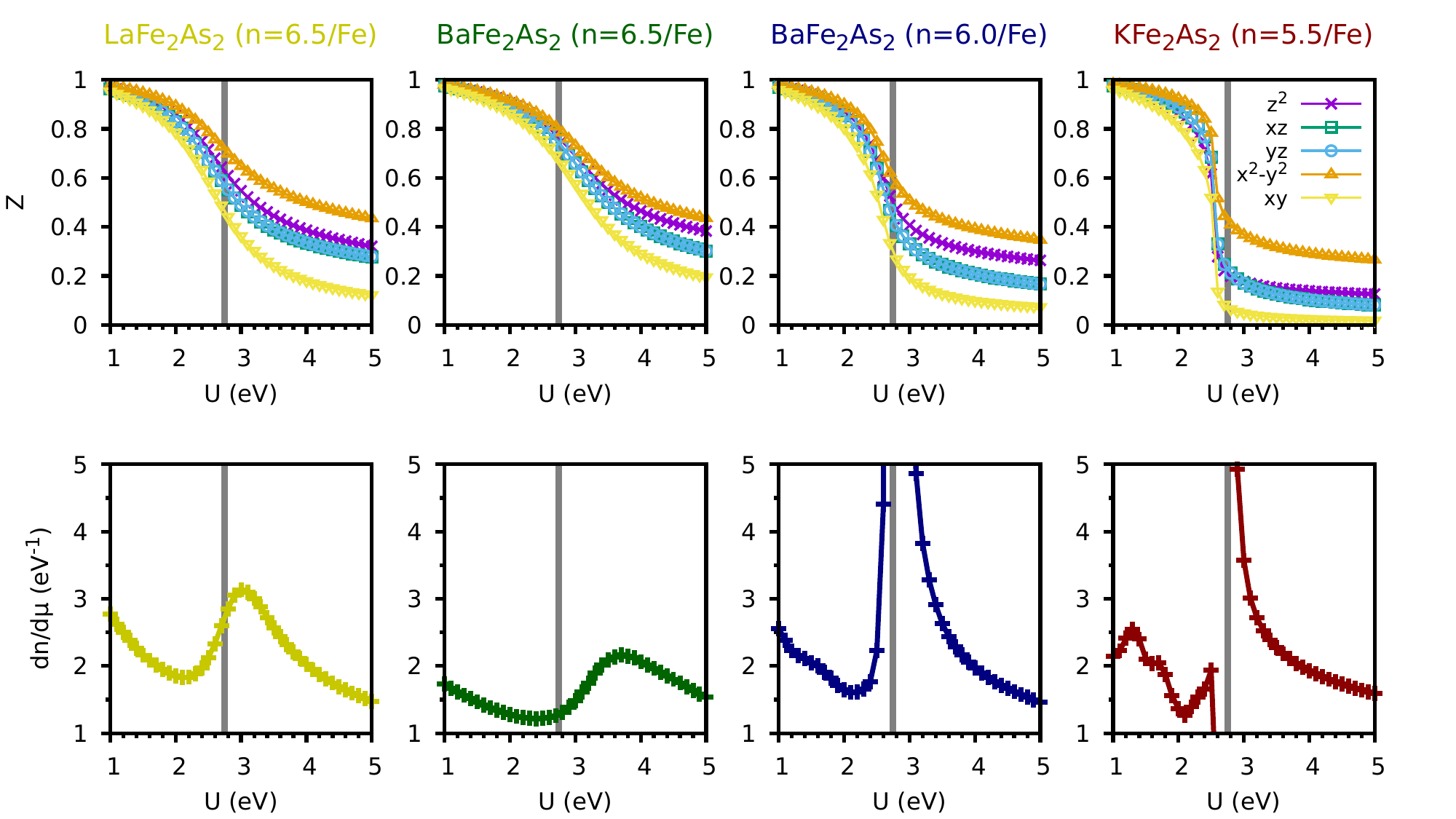}
\caption{
Top panels: quasiparticle weights $Z$ of 122 family computed within the SSMF@PBE approach as a function of the Hubbard $U$.
Top panels: Corresponding electronic compressibility.
The Hund coupling strength has been set by keeping $J/U=0.25$.
\label{fig:Z-and-compressibility}
}
%%%%%%%%%%
\end{figure}

%
% Compressibility/correlation effects
%
Such a finding naturally places LaFe$_2$As$_2$ in the realm of correlated metals, so questioning a pure interpretation of experimental findings in terms of effective doping of an uncorrelated band structure.
In order to place LaFe$_2$As$_2$ within this framework, we report in Fig.~\ref{fig:Z-and-compressibility} the mass renormalizations $Z_m$ as a function of the interaction strength.
As it appears from the top panels, LaFe$_2$As$_2$ displays a smoothened cross-over region similar to electron-overdoped (n=6.5/Fe) BaFe$_2$As$_2$, however, for the same value of the local interaction strength, the former is more correlated than the latter, with effective masses up to twice the PBE ones in the case of the $xy$ orbitals.
Considering that the same interaction strength has been used for both cases, such a behaviour can be explained by the smaller bare bandwidth $W$ of LaFe$_2$As$_2$ with respect to BaFe$_2$As$_2$, so that a bigger $U/W$ ratio is reached.
Our model predicts therefore that LaFe$_2$As$_2$ is not electronically equivalent to BaFe$_2$As$_2$ with $n= 6.5\,e^-$/Fe, but rather displays features which are compatible with an intermediate filling between BaFe$_2$As$_2$(n=6.5/Fe) and BaFe$_2$As$_2$(n= 6.0/Fe). 
It must be stressed that this evolution is entirely due to dynamical correlation effects, and cannot therefore be captured by an independent-band model.

Such a doping evolution of electronic correlations has been shown to be an ubiquitous feature in IBSC, and reconciles several experimental observation within the K$_{x}$Ba$_{1-x}$Fe$_2$As$_2$ family~\cite{deMedici:2014}.
Interestingly, the Hubbard-Kanamori Hamiltonian is believed to display a region of enhanced electronic compressibility ${\rm d}n/{\rm d}\mu$ in its $(n,U)$ phase diagram, which separates the normal metal from the Hund metal region~\cite{deMedici:2017ab,Chatzieleftheriou:2020}.
%~\cite{deMedici:2009,deMedici:2011,Yin:2011,Si:2012,Georges:2013,deMedici:2017}.
%
It has also been recently speculated that the position of a material with respect to such an enhanced-compressibility region may be related to its superconducting critical temperature~\cite{Villar-Arribi:2018}.
Inspection of the electronic compressibility, reported in the lower panels of Fig.~\ref{fig:Z-and-compressibility}, reveals that LaFe$_2$As$_2$ results closer to the enhancement region than electron-doped BaFe$_2$As$_2$, further confirming the stronger degree of electronic correlations in the former, and possibly suggesting a key role of electronic correrations in the building of the the IBSC superconducting state.

%
% \cite{Acharya:2020} --> DMFT@QSGW
%
% U = 3.5 eV, J = 0.62 eV (J/U= 0.177)
% quasiparticle weights Z and scattering rates (meV)
%  z2    = 0.5      32
% xz/yz = 0.54    60
% x2-y2 = 0.54   28
% xy      = 0.49   67
%
% cRPA calculations report U = 3.88 and J = 0.72 (J/U=0.186)
%
% pd model?

%%%%%%%%%%%%%%
%
\section{Conclusions}
%
%%%%%%%%%%%%%%

In conclusion, we have studied the electronic properties of LaFe$_2$As$_2$ by including electronic correlations in a realistic fashion by means of SSMF@DFT calculations.
By leveraging on the numerical agility of our approach, we could freely explore the local interaction space and found that the experimental Sommerfeld coefficient is reproduced by $U=2.7\div2.8$\,eV and $J/U=0.25$, the same values capturing its doping evolution across the K$_{x}$Ba$_{1-x}$Fe$_2$As$_2$ series.
A comparison of the computed electronic properties between LaFe$_2$As$_2$ and the K$_{x}$Ba$_{1-x}$Fe$_2$As$_2$ series reveals that electronic correlations are stronger in LaFe$_2$As$_2$ for the same nominal filling of 6.5 electrons/Fe, due to the smaller bare bandwidth of the former.
Our result points at the strength of electronic correlations, rather than sheer doping, as the key parameter in assessing the IBSC phase diagram.

%%%%%%%%%%%%%%%%%
\section*{ACKNOWLEDGMENTS}
%%%%%%%%%%%%%%%%%
The authors acknowledge fruitful discussions with I. Pallecchi, M. Putti and I.\ Timrov.
TG and LdM are supported by the European Commission through the ERC-CoG2016, StrongCoPhy4Energy, GA No724177.
This work was granted access to the HPC resources of MesoPSL financed by the Region \^Ile-de-France and the project Equip@Meso (reference ANR-10-EQPX-29-01) of the programme Investissements d'Avenir supervised by the Agence Nationale pour la Recherche.

%%%%%%%%%%
\onecolumngrid

\twocolumngrid
%%%%%%%%%%

%%%%%%%%%%%%%%%%%%%%%%%%%%
\bibliography{IBSC,DFT,DMFTrealsystems,Hund}

\begin{thebibliography}{41}%
\makeatletter
\providecommand \@ifxundefined [1]{%
 \@ifx{#1\undefined}
}%
\providecommand \@ifnum [1]{%
 \ifnum #1\expandafter \@firstoftwo
 \else \expandafter \@secondoftwo
 \fi
}%
\providecommand \@ifx [1]{%
 \ifx #1\expandafter \@firstoftwo
 \else \expandafter \@secondoftwo
 \fi
}%
\providecommand \natexlab [1]{#1}%
\providecommand \enquote  [1]{``#1''}%
\providecommand \bibnamefont  [1]{#1}%
\providecommand \bibfnamefont [1]{#1}%
\providecommand \citenamefont [1]{#1}%
\providecommand \href@noop [0]{\@secondoftwo}%
\providecommand \href [0]{\begingroup \@sanitize@url \@href}%
\providecommand \@href[1]{\@@startlink{#1}\@@href}%
\providecommand \@@href[1]{\endgroup#1\@@endlink}%
\providecommand \@sanitize@url [0]{\catcode `\\12\catcode `\$12\catcode
  `\&12\catcode `\#12\catcode `\^12\catcode `\_12\catcode `\%12\relax}%
\providecommand \@@startlink[1]{}%
\providecommand \@@endlink[0]{}%
\providecommand \url  [0]{\begingroup\@sanitize@url \@url }%
\providecommand \@url [1]{\endgroup\@href {#1}{\urlprefix }}%
\providecommand \urlprefix  [0]{URL }%
\providecommand \Eprint [0]{\href }%
\providecommand \doibase [0]{http://dx.doi.org/}%
\providecommand \selectlanguage [0]{\@gobble}%
\providecommand \bibinfo  [0]{\@secondoftwo}%
\providecommand \bibfield  [0]{\@secondoftwo}%
\providecommand \translation [1]{[#1]}%
\providecommand \BibitemOpen [0]{}%
\providecommand \bibitemStop [0]{}%
\providecommand \bibitemNoStop [0]{.\EOS\space}%
\providecommand \EOS [0]{\spacefactor3000\relax}%
\providecommand \BibitemShut  [1]{\csname bibitem#1\endcsname}%
\let\auto@bib@innerbib\@empty
%</preamble>
\bibitem [{\citenamefont {Johnston}(2010)}]{revJohnston:2010}%
  \BibitemOpen
  \bibfield  {author} {\bibinfo {author} {\bibfnamefont {D.~C.}\ \bibnamefont
  {Johnston}},\ }\href@noop {} {\bibfield  {journal} {\bibinfo  {journal}
  {Advances in Physics}\ }\textbf {\bibinfo {volume} {59}},\ \bibinfo {pages}
  {803} (\bibinfo {year} {2010})}\BibitemShut {NoStop}%
\bibitem [{\citenamefont {Johnson}\ \emph {et~al.}(2015)\citenamefont
  {Johnson}, \citenamefont {Xu},\ and\ \citenamefont {Yin}}]{revJohnson:2015}%
  \BibitemOpen
  \bibfield  {author} {\bibinfo {author} {\bibfnamefont {P.~D.}\ \bibnamefont
  {Johnson}}, \bibinfo {author} {\bibfnamefont {G.}~\bibnamefont {Xu}}, \ and\
  \bibinfo {author} {\bibfnamefont {W.-G.}\ \bibnamefont {Yin}},\ }\href@noop
  {} {\emph {\bibinfo {title} {Iron-based superconductivity}}},\ Vol.\ \bibinfo
  {volume} {211}\ (\bibinfo  {publisher} {Springer},\ \bibinfo {year}
  {2015})\BibitemShut {NoStop}%
\bibitem [{\citenamefont {Martinelli}\ \emph {et~al.}(2016)\citenamefont
  {Martinelli}, \citenamefont {Bernardini},\ and\ \citenamefont
  {Massidda}}]{revMartinelli:2016}%
  \BibitemOpen
  \bibfield  {author} {\bibinfo {author} {\bibfnamefont {A.}~\bibnamefont
  {Martinelli}}, \bibinfo {author} {\bibfnamefont {F.}~\bibnamefont
  {Bernardini}}, \ and\ \bibinfo {author} {\bibfnamefont {S.}~\bibnamefont
  {Massidda}},\ }\href@noop {} {\bibfield  {journal} {\bibinfo  {journal}
  {Comptes Rendus Physique}\ }\textbf {\bibinfo {volume} {17}},\ \bibinfo
  {pages} {5} (\bibinfo {year} {2016})}\BibitemShut {NoStop}%
\bibitem [{\citenamefont {Stewart}(2011)}]{revStewart:2011}%
  \BibitemOpen
  \bibfield  {author} {\bibinfo {author} {\bibfnamefont {G.}~\bibnamefont
  {Stewart}},\ }\href@noop {} {\bibfield  {journal} {\bibinfo  {journal}
  {Reviews of Modern Physics}\ }\textbf {\bibinfo {volume} {83}},\ \bibinfo
  {pages} {1589} (\bibinfo {year} {2011})}\BibitemShut {NoStop}%
\bibitem [{\citenamefont {Dai}(2015)}]{Dai:2015}%
  \BibitemOpen
  \bibfield  {author} {\bibinfo {author} {\bibfnamefont {P.}~\bibnamefont
  {Dai}},\ }\href {\doibase 10.1103/RevModPhys.87.855} {\bibfield  {journal}
  {\bibinfo  {journal} {Rev. Mod. Phys.}\ }\textbf {\bibinfo {volume} {87}},\
  \bibinfo {pages} {855} (\bibinfo {year} {2015})}\BibitemShut {NoStop}%
\bibitem [{\citenamefont {Hardy}\ \emph {et~al.}(2010)\citenamefont {Hardy},
  \citenamefont {Burger}, \citenamefont {Wolf}, \citenamefont {Fisher},
  \citenamefont {Schweiss}, \citenamefont {Adelmann}, \citenamefont {Heid},
  \citenamefont {Fromknecht}, \citenamefont {Eder}, \citenamefont {Ernst} \emph
  {et~al.}}]{Hardy:2010}%
  \BibitemOpen
  \bibfield  {author} {\bibinfo {author} {\bibfnamefont {F.}~\bibnamefont
  {Hardy}}, \bibinfo {author} {\bibfnamefont {P.}~\bibnamefont {Burger}},
  \bibinfo {author} {\bibfnamefont {T.}~\bibnamefont {Wolf}}, \bibinfo {author}
  {\bibfnamefont {R.}~\bibnamefont {Fisher}}, \bibinfo {author} {\bibfnamefont
  {P.}~\bibnamefont {Schweiss}}, \bibinfo {author} {\bibfnamefont
  {P.}~\bibnamefont {Adelmann}}, \bibinfo {author} {\bibfnamefont
  {R.}~\bibnamefont {Heid}}, \bibinfo {author} {\bibfnamefont {R.}~\bibnamefont
  {Fromknecht}}, \bibinfo {author} {\bibfnamefont {R.}~\bibnamefont {Eder}},
  \bibinfo {author} {\bibfnamefont {D.}~\bibnamefont {Ernst}},  \emph
  {et~al.},\ }\href@noop {} {\bibfield  {journal} {\bibinfo  {journal} {EPL
  (Europhysics Letters)}\ }\textbf {\bibinfo {volume} {91}},\ \bibinfo {pages}
  {47008} (\bibinfo {year} {2010})}\BibitemShut {NoStop}%
\bibitem [{\citenamefont {Hardy}\ \emph {et~al.}(2016)\citenamefont {Hardy},
  \citenamefont {B{\"o}hmer}, \citenamefont {de' Medici}, \citenamefont
  {Capone}, \citenamefont {Giovannetti}, \citenamefont {Eder}, \citenamefont
  {Wang}, \citenamefont {He}, \citenamefont {Wolf}, \citenamefont {Schweiss}
  \emph {et~al.}}]{Hardy:2016}%
  \BibitemOpen
  \bibfield  {author} {\bibinfo {author} {\bibfnamefont {F.}~\bibnamefont
  {Hardy}}, \bibinfo {author} {\bibfnamefont {A.}~\bibnamefont {B{\"o}hmer}},
  \bibinfo {author} {\bibfnamefont {L.}~\bibnamefont {de' Medici}}, \bibinfo
  {author} {\bibfnamefont {M.}~\bibnamefont {Capone}}, \bibinfo {author}
  {\bibfnamefont {G.}~\bibnamefont {Giovannetti}}, \bibinfo {author}
  {\bibfnamefont {R.}~\bibnamefont {Eder}}, \bibinfo {author} {\bibfnamefont
  {L.}~\bibnamefont {Wang}}, \bibinfo {author} {\bibfnamefont {M.}~\bibnamefont
  {He}}, \bibinfo {author} {\bibfnamefont {T.}~\bibnamefont {Wolf}}, \bibinfo
  {author} {\bibfnamefont {P.}~\bibnamefont {Schweiss}},  \emph {et~al.},\
  }\href@noop {} {\bibfield  {journal} {\bibinfo  {journal} {Physical Review
  B}\ }\textbf {\bibinfo {volume} {94}},\ \bibinfo {pages} {205113} (\bibinfo
  {year} {2016})}\BibitemShut {NoStop}%
\bibitem [{\citenamefont {Iyo}\ \emph {et~al.}(2018)\citenamefont {Iyo},
  \citenamefont {Kawashima}, \citenamefont {Ishida}, \citenamefont {Fujihisa},
  \citenamefont {Gotoh}, \citenamefont {Eisaki},\ and\ \citenamefont
  {Yoshida}}]{Iyo:2018}%
  \BibitemOpen
  \bibfield  {author} {\bibinfo {author} {\bibfnamefont {A.}~\bibnamefont
  {Iyo}}, \bibinfo {author} {\bibfnamefont {K.}~\bibnamefont {Kawashima}},
  \bibinfo {author} {\bibfnamefont {S.}~\bibnamefont {Ishida}}, \bibinfo
  {author} {\bibfnamefont {H.}~\bibnamefont {Fujihisa}}, \bibinfo {author}
  {\bibfnamefont {Y.}~\bibnamefont {Gotoh}}, \bibinfo {author} {\bibfnamefont
  {H.}~\bibnamefont {Eisaki}}, \ and\ \bibinfo {author} {\bibfnamefont
  {Y.}~\bibnamefont {Yoshida}},\ }\href@noop {} {\bibfield  {journal} {\bibinfo
   {journal} {Journal of the American Chemical Society}\ }\textbf {\bibinfo
  {volume} {140}},\ \bibinfo {pages} {369} (\bibinfo {year}
  {2018})}\BibitemShut {NoStop}%
\bibitem [{\citenamefont {Iyo}\ \emph {et~al.}(2019)\citenamefont {Iyo},
  \citenamefont {Ishida}, \citenamefont {Fujihisa}, \citenamefont {Gotoh},
  \citenamefont {Hase}, \citenamefont {Yoshida}, \citenamefont {Eisaki},\ and\
  \citenamefont {Kawashima}}]{Iyo:2019}%
  \BibitemOpen
  \bibfield  {author} {\bibinfo {author} {\bibfnamefont {A.}~\bibnamefont
  {Iyo}}, \bibinfo {author} {\bibfnamefont {S.}~\bibnamefont {Ishida}},
  \bibinfo {author} {\bibfnamefont {H.}~\bibnamefont {Fujihisa}}, \bibinfo
  {author} {\bibfnamefont {Y.}~\bibnamefont {Gotoh}}, \bibinfo {author}
  {\bibfnamefont {I.}~\bibnamefont {Hase}}, \bibinfo {author} {\bibfnamefont
  {Y.}~\bibnamefont {Yoshida}}, \bibinfo {author} {\bibfnamefont
  {H.}~\bibnamefont {Eisaki}}, \ and\ \bibinfo {author} {\bibfnamefont
  {K.}~\bibnamefont {Kawashima}},\ }\href@noop {} {\bibfield  {journal}
  {\bibinfo  {journal} {The journal of physical chemistry letters}\ }\textbf
  {\bibinfo {volume} {10}},\ \bibinfo {pages} {1018} (\bibinfo {year}
  {2019})}\BibitemShut {NoStop}%
\bibitem [{\citenamefont {Usui}\ and\ \citenamefont
  {Kuroki}(2019)}]{Usui:2019}%
  \BibitemOpen
  \bibfield  {author} {\bibinfo {author} {\bibfnamefont {H.}~\bibnamefont
  {Usui}}\ and\ \bibinfo {author} {\bibfnamefont {K.}~\bibnamefont {Kuroki}},\
  }\href {\doibase 10.1103/PhysRevResearch.1.033025} {\bibfield  {journal}
  {\bibinfo  {journal} {Phys. Rev. Research}\ }\textbf {\bibinfo {volume}
  {1}},\ \bibinfo {pages} {033025} (\bibinfo {year} {2019})}\BibitemShut
  {NoStop}%
\bibitem [{\citenamefont {Mazin}\ \emph {et~al.}(2019)\citenamefont {Mazin},
  \citenamefont {Shimizu}, \citenamefont {Takemori},\ and\ \citenamefont
  {Jeschke}}]{Mazin:2019}%
  \BibitemOpen
  \bibfield  {author} {\bibinfo {author} {\bibfnamefont {I.~I.}\ \bibnamefont
  {Mazin}}, \bibinfo {author} {\bibfnamefont {M.}~\bibnamefont {Shimizu}},
  \bibinfo {author} {\bibfnamefont {N.}~\bibnamefont {Takemori}}, \ and\
  \bibinfo {author} {\bibfnamefont {H.~O.}\ \bibnamefont {Jeschke}},\ }\href
  {\doibase 10.1103/PhysRevLett.123.267001} {\bibfield  {journal} {\bibinfo
  {journal} {Phys. Rev. Lett.}\ }\textbf {\bibinfo {volume} {123}},\ \bibinfo
  {pages} {267001} (\bibinfo {year} {2019})}\BibitemShut {NoStop}%
\bibitem [{\citenamefont {Coldea}\ \emph {et~al.}(2008)\citenamefont {Coldea},
  \citenamefont {Fletcher}, \citenamefont {Carrington}, \citenamefont
  {Analytis}, \citenamefont {Bangura}, \citenamefont {Chu}, \citenamefont
  {Erickson}, \citenamefont {Fisher}, \citenamefont {Hussey},\ and\
  \citenamefont {McDonald}}]{Coldea:2008}%
  \BibitemOpen
  \bibfield  {author} {\bibinfo {author} {\bibfnamefont {A.}~\bibnamefont
  {Coldea}}, \bibinfo {author} {\bibfnamefont {J.}~\bibnamefont {Fletcher}},
  \bibinfo {author} {\bibfnamefont {A.}~\bibnamefont {Carrington}}, \bibinfo
  {author} {\bibfnamefont {J.}~\bibnamefont {Analytis}}, \bibinfo {author}
  {\bibfnamefont {A.}~\bibnamefont {Bangura}}, \bibinfo {author} {\bibfnamefont
  {J.-H.}\ \bibnamefont {Chu}}, \bibinfo {author} {\bibfnamefont
  {A.}~\bibnamefont {Erickson}}, \bibinfo {author} {\bibfnamefont
  {I.}~\bibnamefont {Fisher}}, \bibinfo {author} {\bibfnamefont
  {N.}~\bibnamefont {Hussey}}, \ and\ \bibinfo {author} {\bibfnamefont
  {R.}~\bibnamefont {McDonald}},\ }\href@noop {} {\bibfield  {journal}
  {\bibinfo  {journal} {Physical review letters}\ }\textbf {\bibinfo {volume}
  {101}},\ \bibinfo {pages} {216402} (\bibinfo {year} {2008})}\BibitemShut
  {NoStop}%
\bibitem [{\citenamefont {Brouet}\ \emph {et~al.}(2009)\citenamefont {Brouet},
  \citenamefont {Marsi}, \citenamefont {Mansart}, \citenamefont {Nicolaou},
  \citenamefont {Taleb-Ibrahimi}, \citenamefont {Le~Fevre}, \citenamefont
  {Bertran}, \citenamefont {Rullier-Albenque}, \citenamefont {Forget},\ and\
  \citenamefont {Colson}}]{Brouet:2009}%
  \BibitemOpen
  \bibfield  {author} {\bibinfo {author} {\bibfnamefont {V.}~\bibnamefont
  {Brouet}}, \bibinfo {author} {\bibfnamefont {M.}~\bibnamefont {Marsi}},
  \bibinfo {author} {\bibfnamefont {B.}~\bibnamefont {Mansart}}, \bibinfo
  {author} {\bibfnamefont {A.}~\bibnamefont {Nicolaou}}, \bibinfo {author}
  {\bibfnamefont {A.}~\bibnamefont {Taleb-Ibrahimi}}, \bibinfo {author}
  {\bibfnamefont {P.}~\bibnamefont {Le~Fevre}}, \bibinfo {author}
  {\bibfnamefont {F.}~\bibnamefont {Bertran}}, \bibinfo {author} {\bibfnamefont
  {F.}~\bibnamefont {Rullier-Albenque}}, \bibinfo {author} {\bibfnamefont
  {A.}~\bibnamefont {Forget}}, \ and\ \bibinfo {author} {\bibfnamefont
  {D.}~\bibnamefont {Colson}},\ }\href@noop {} {\bibfield  {journal} {\bibinfo
  {journal} {Physical review B}\ }\textbf {\bibinfo {volume} {80}},\ \bibinfo
  {pages} {165115} (\bibinfo {year} {2009})}\BibitemShut {NoStop}%
\bibitem [{\citenamefont {Lee}\ \emph {et~al.}(2012)\citenamefont {Lee},
  \citenamefont {Ji}, \citenamefont {Kim}, \citenamefont {Kim}, \citenamefont
  {Haule}, \citenamefont {Kotliar}, \citenamefont {Lee}, \citenamefont {Khim},
  \citenamefont {Kim}, \citenamefont {Kim} \emph {et~al.}}]{Lee:2012}%
  \BibitemOpen
  \bibfield  {author} {\bibinfo {author} {\bibfnamefont {G.}~\bibnamefont
  {Lee}}, \bibinfo {author} {\bibfnamefont {H.~S.}\ \bibnamefont {Ji}},
  \bibinfo {author} {\bibfnamefont {Y.}~\bibnamefont {Kim}}, \bibinfo {author}
  {\bibfnamefont {C.}~\bibnamefont {Kim}}, \bibinfo {author} {\bibfnamefont
  {K.}~\bibnamefont {Haule}}, \bibinfo {author} {\bibfnamefont
  {G.}~\bibnamefont {Kotliar}}, \bibinfo {author} {\bibfnamefont
  {B.}~\bibnamefont {Lee}}, \bibinfo {author} {\bibfnamefont {S.}~\bibnamefont
  {Khim}}, \bibinfo {author} {\bibfnamefont {K.~H.}\ \bibnamefont {Kim}},
  \bibinfo {author} {\bibfnamefont {K.~S.}\ \bibnamefont {Kim}},  \emph
  {et~al.},\ }\href@noop {} {\bibfield  {journal} {\bibinfo  {journal}
  {Physical review letters}\ }\textbf {\bibinfo {volume} {109}},\ \bibinfo
  {pages} {177001} (\bibinfo {year} {2012})}\BibitemShut {NoStop}%
\bibitem [{\citenamefont {Terashima}\ \emph {et~al.}(2013)\citenamefont
  {Terashima}, \citenamefont {Kurita}, \citenamefont {Kimata}, \citenamefont
  {Tomita}, \citenamefont {Tsuchiya}, \citenamefont {Satsukawa}, \citenamefont
  {Harada}, \citenamefont {Hazama}, \citenamefont {Imai}, \citenamefont {Sato}
  \emph {et~al.}}]{Terashima:2013}%
  \BibitemOpen
  \bibfield  {author} {\bibinfo {author} {\bibfnamefont {T.}~\bibnamefont
  {Terashima}}, \bibinfo {author} {\bibfnamefont {N.}~\bibnamefont {Kurita}},
  \bibinfo {author} {\bibfnamefont {M.}~\bibnamefont {Kimata}}, \bibinfo
  {author} {\bibfnamefont {M.}~\bibnamefont {Tomita}}, \bibinfo {author}
  {\bibfnamefont {S.}~\bibnamefont {Tsuchiya}}, \bibinfo {author}
  {\bibfnamefont {H.}~\bibnamefont {Satsukawa}}, \bibinfo {author}
  {\bibfnamefont {A.}~\bibnamefont {Harada}}, \bibinfo {author} {\bibfnamefont
  {K.}~\bibnamefont {Hazama}}, \bibinfo {author} {\bibfnamefont
  {M.}~\bibnamefont {Imai}}, \bibinfo {author} {\bibfnamefont {A.}~\bibnamefont
  {Sato}},  \emph {et~al.},\ }in\ \href@noop {} {\emph {\bibinfo {booktitle}
  {Journal of Physics: Conference Series}}},\ Vol.\ \bibinfo {volume} {449}\
  (\bibinfo {organization} {IOP Publishing},\ \bibinfo {year} {2013})\ p.\
  \bibinfo {pages} {012022}\BibitemShut {NoStop}%
\bibitem [{\citenamefont {Watson}\ \emph {et~al.}(2015)\citenamefont {Watson},
  \citenamefont {Kim}, \citenamefont {Haghighirad}, \citenamefont {Davies},
  \citenamefont {McCollam}, \citenamefont {Narayanan}, \citenamefont {Blake},
  \citenamefont {Chen}, \citenamefont {Ghannadzadeh}, \citenamefont
  {Schofield}, \citenamefont {Hoesch}, \citenamefont {Meingast}, \citenamefont
  {Wolf},\ and\ \citenamefont {Coldea}}]{Watson:2015}%
  \BibitemOpen
  \bibfield  {author} {\bibinfo {author} {\bibfnamefont {M.~D.}\ \bibnamefont
  {Watson}}, \bibinfo {author} {\bibfnamefont {T.~K.}\ \bibnamefont {Kim}},
  \bibinfo {author} {\bibfnamefont {A.~A.}\ \bibnamefont {Haghighirad}},
  \bibinfo {author} {\bibfnamefont {N.~R.}\ \bibnamefont {Davies}}, \bibinfo
  {author} {\bibfnamefont {A.}~\bibnamefont {McCollam}}, \bibinfo {author}
  {\bibfnamefont {A.}~\bibnamefont {Narayanan}}, \bibinfo {author}
  {\bibfnamefont {S.~F.}\ \bibnamefont {Blake}}, \bibinfo {author}
  {\bibfnamefont {Y.~L.}\ \bibnamefont {Chen}}, \bibinfo {author}
  {\bibfnamefont {S.}~\bibnamefont {Ghannadzadeh}}, \bibinfo {author}
  {\bibfnamefont {A.~J.}\ \bibnamefont {Schofield}}, \bibinfo {author}
  {\bibfnamefont {M.}~\bibnamefont {Hoesch}}, \bibinfo {author} {\bibfnamefont
  {C.}~\bibnamefont {Meingast}}, \bibinfo {author} {\bibfnamefont
  {T.}~\bibnamefont {Wolf}}, \ and\ \bibinfo {author} {\bibfnamefont {A.~I.}\
  \bibnamefont {Coldea}},\ }\href {\doibase 10.1103/PhysRevB.91.155106}
  {\bibfield  {journal} {\bibinfo  {journal} {Phys. Rev. B}\ }\textbf {\bibinfo
  {volume} {91}},\ \bibinfo {pages} {155106} (\bibinfo {year}
  {2015})}\BibitemShut {NoStop}%
\bibitem [{\citenamefont {Kouchi}\ \emph {et~al.}(2019)\citenamefont {Kouchi},
  \citenamefont {Yashima}, \citenamefont {Mukuda}, \citenamefont {Ishida},
  \citenamefont {Eisaki}, \citenamefont {Yoshida}, \citenamefont {Kawashima},\
  and\ \citenamefont {Iyo}}]{Kouchi:2019}%
  \BibitemOpen
  \bibfield  {author} {\bibinfo {author} {\bibfnamefont {T.}~\bibnamefont
  {Kouchi}}, \bibinfo {author} {\bibfnamefont {M.}~\bibnamefont {Yashima}},
  \bibinfo {author} {\bibfnamefont {H.}~\bibnamefont {Mukuda}}, \bibinfo
  {author} {\bibfnamefont {S.}~\bibnamefont {Ishida}}, \bibinfo {author}
  {\bibfnamefont {H.}~\bibnamefont {Eisaki}}, \bibinfo {author} {\bibfnamefont
  {Y.}~\bibnamefont {Yoshida}}, \bibinfo {author} {\bibfnamefont
  {K.}~\bibnamefont {Kawashima}}, \ and\ \bibinfo {author} {\bibfnamefont
  {A.}~\bibnamefont {Iyo}},\ }\href@noop {} {\bibfield  {journal} {\bibinfo
  {journal} {Journal of the Physical Society of Japan}\ }\textbf {\bibinfo
  {volume} {88}},\ \bibinfo {pages} {113702} (\bibinfo {year}
  {2019})}\BibitemShut {NoStop}%
\bibitem [{\citenamefont {Pallecchi}\ \emph {et~al.}(2020)\citenamefont
  {Pallecchi}, \citenamefont {Iyo}, \citenamefont {Ogino}, \citenamefont
  {Affronte},\ and\ \citenamefont {Putti}}]{Pallecchi:2020}%
  \BibitemOpen
  \bibfield  {author} {\bibinfo {author} {\bibfnamefont {I.}~\bibnamefont
  {Pallecchi}}, \bibinfo {author} {\bibfnamefont {A.}~\bibnamefont {Iyo}},
  \bibinfo {author} {\bibfnamefont {H.}~\bibnamefont {Ogino}}, \bibinfo
  {author} {\bibfnamefont {M.}~\bibnamefont {Affronte}}, \ and\ \bibinfo
  {author} {\bibfnamefont {M.}~\bibnamefont {Putti}},\ }\href {\doibase
  10.1103/PhysRevMaterials.4.114803} {\bibfield  {journal} {\bibinfo  {journal}
  {Phys. Rev. Materials}\ }\textbf {\bibinfo {volume} {4}},\ \bibinfo {pages}
  {114803} (\bibinfo {year} {2020})}\BibitemShut {NoStop}%
\bibitem [{\citenamefont {Yin}\ \emph {et~al.}(2011)\citenamefont {Yin},
  \citenamefont {Haule},\ and\ \citenamefont {Kotliar}}]{Yin:2011}%
  \BibitemOpen
  \bibfield  {author} {\bibinfo {author} {\bibfnamefont {Z.}~\bibnamefont
  {Yin}}, \bibinfo {author} {\bibfnamefont {K.}~\bibnamefont {Haule}}, \ and\
  \bibinfo {author} {\bibfnamefont {G.}~\bibnamefont {Kotliar}},\ }\href@noop
  {} {\bibfield  {journal} {\bibinfo  {journal} {Nature materials}\ }\textbf
  {\bibinfo {volume} {10}},\ \bibinfo {pages} {932} (\bibinfo {year}
  {2011})}\BibitemShut {NoStop}%
\bibitem [{\citenamefont {de' Medici}\ \emph {et~al.}(2014)\citenamefont {de'
  Medici}, \citenamefont {Giovannetti},\ and\ \citenamefont
  {Capone}}]{deMedici:2014}%
  \BibitemOpen
  \bibfield  {author} {\bibinfo {author} {\bibfnamefont {L.}~\bibnamefont {de'
  Medici}}, \bibinfo {author} {\bibfnamefont {G.}~\bibnamefont {Giovannetti}},
  \ and\ \bibinfo {author} {\bibfnamefont {M.}~\bibnamefont {Capone}},\
  }\href@noop {} {\bibfield  {journal} {\bibinfo  {journal} {Physical review
  letters}\ }\textbf {\bibinfo {volume} {112}},\ \bibinfo {pages} {177001}
  (\bibinfo {year} {2014})}\BibitemShut {NoStop}%
\bibitem [{\citenamefont {Gorni}\ \emph {et~al.}(2021)\citenamefont {Gorni},
  \citenamefont {Arribi}, \citenamefont {Casula},\ and\ \citenamefont {de'
  Medici}}]{Gorni:2021}%
  \BibitemOpen
  \bibfield  {author} {\bibinfo {author} {\bibfnamefont {T.}~\bibnamefont
  {Gorni}}, \bibinfo {author} {\bibfnamefont {P.~V.}\ \bibnamefont {Arribi}},
  \bibinfo {author} {\bibfnamefont {M.}~\bibnamefont {Casula}}, \ and\ \bibinfo
  {author} {\bibfnamefont {L.}~\bibnamefont {de' Medici}},\ }\href@noop {}
  {\bibfield  {journal} {\bibinfo  {journal} {Physical Review B}\ }\textbf
  {\bibinfo {volume} {104}},\ \bibinfo {pages} {014507} (\bibinfo {year}
  {2021})}\BibitemShut {NoStop}%
\bibitem [{\citenamefont {Acharya}\ \emph {et~al.}(2020)\citenamefont
  {Acharya}, \citenamefont {Pashov}, \citenamefont {Jamet},\ and\ \citenamefont
  {van Schilfgaarde}}]{Acharya:2020}%
  \BibitemOpen
  \bibfield  {author} {\bibinfo {author} {\bibfnamefont {S.}~\bibnamefont
  {Acharya}}, \bibinfo {author} {\bibfnamefont {D.}~\bibnamefont {Pashov}},
  \bibinfo {author} {\bibfnamefont {F.}~\bibnamefont {Jamet}}, \ and\ \bibinfo
  {author} {\bibfnamefont {M.}~\bibnamefont {van Schilfgaarde}},\ }\href
  {\doibase 10.1103/PhysRevLett.124.237001} {\bibfield  {journal} {\bibinfo
  {journal} {Phys. Rev. Lett.}\ }\textbf {\bibinfo {volume} {124}},\ \bibinfo
  {pages} {237001} (\bibinfo {year} {2020})}\BibitemShut {NoStop}%
\bibitem [{\citenamefont {Zhao}\ \emph {et~al.}(2021)\citenamefont {Zhao},
  \citenamefont {Wang}, \citenamefont {Feng},\ and\ \citenamefont
  {Yang}}]{Zhao:2021}%
  \BibitemOpen
  \bibfield  {author} {\bibinfo {author} {\bibfnamefont {J.}~\bibnamefont
  {Zhao}}, \bibinfo {author} {\bibfnamefont {Y.}~\bibnamefont {Wang}}, \bibinfo
  {author} {\bibfnamefont {X.}~\bibnamefont {Feng}}, \ and\ \bibinfo {author}
  {\bibfnamefont {S.~A.}\ \bibnamefont {Yang}},\ }\href {\doibase
  10.1103/PhysRevB.103.125125} {\bibfield  {journal} {\bibinfo  {journal}
  {Phys. Rev. B}\ }\textbf {\bibinfo {volume} {103}},\ \bibinfo {pages}
  {125125} (\bibinfo {year} {2021})}\BibitemShut {NoStop}%
\bibitem [{\citenamefont {Giannozzi}\ \emph {et~al.}(2009)\citenamefont
  {Giannozzi}, \citenamefont {Baroni}, \citenamefont {Bonini}, \citenamefont
  {Calandra}, \citenamefont {Car}, \citenamefont {Cavazzoni}, \citenamefont
  {Ceresoli}, \citenamefont {Chiarotti}, \citenamefont {Cococcioni},
  \citenamefont {Dabo}, \citenamefont {Dal~Corso}, \citenamefont
  {De~Gironcoli}, \citenamefont {Fabris}, \citenamefont {Fratesi},
  \citenamefont {Gebauer}, \citenamefont {Gerstmann}, \citenamefont
  {Gougoussis}, \citenamefont {Kokalj}, \citenamefont {Lazzeri}, \citenamefont
  {Martin-Samos}, \citenamefont {Marzari}, \citenamefont {Mauri}, \citenamefont
  {Mazzarello}, \citenamefont {Paolini}, \citenamefont {Pasquarello},
  \citenamefont {Paulatto}, \citenamefont {Sbraccia}, \citenamefont {Scandolo},
  \citenamefont {Sclauzero}, \citenamefont {Seitsonen}, \citenamefont
  {Smogunov}, \citenamefont {Umari},\ and\ \citenamefont
  {Wentzcovitch}}]{Giannozzi:2009}%
  \BibitemOpen
  \bibfield  {author} {\bibinfo {author} {\bibfnamefont {P.}~\bibnamefont
  {Giannozzi}}, \bibinfo {author} {\bibfnamefont {S.}~\bibnamefont {Baroni}},
  \bibinfo {author} {\bibfnamefont {N.}~\bibnamefont {Bonini}}, \bibinfo
  {author} {\bibfnamefont {M.}~\bibnamefont {Calandra}}, \bibinfo {author}
  {\bibfnamefont {R.}~\bibnamefont {Car}}, \bibinfo {author} {\bibfnamefont
  {C.}~\bibnamefont {Cavazzoni}}, \bibinfo {author} {\bibfnamefont
  {D.}~\bibnamefont {Ceresoli}}, \bibinfo {author} {\bibfnamefont
  {G.}~\bibnamefont {Chiarotti}}, \bibinfo {author} {\bibfnamefont
  {M.}~\bibnamefont {Cococcioni}}, \bibinfo {author} {\bibfnamefont
  {I.}~\bibnamefont {Dabo}}, \bibinfo {author} {\bibfnamefont {A.}~\bibnamefont
  {Dal~Corso}}, \bibinfo {author} {\bibfnamefont {S.}~\bibnamefont
  {De~Gironcoli}}, \bibinfo {author} {\bibfnamefont {S.}~\bibnamefont
  {Fabris}}, \bibinfo {author} {\bibfnamefont {G.}~\bibnamefont {Fratesi}},
  \bibinfo {author} {\bibfnamefont {R.}~\bibnamefont {Gebauer}}, \bibinfo
  {author} {\bibfnamefont {U.}~\bibnamefont {Gerstmann}}, \bibinfo {author}
  {\bibfnamefont {C.}~\bibnamefont {Gougoussis}}, \bibinfo {author}
  {\bibfnamefont {A.}~\bibnamefont {Kokalj}}, \bibinfo {author} {\bibfnamefont
  {M.}~\bibnamefont {Lazzeri}}, \bibinfo {author} {\bibfnamefont
  {L.}~\bibnamefont {Martin-Samos}}, \bibinfo {author} {\bibfnamefont
  {N.}~\bibnamefont {Marzari}}, \bibinfo {author} {\bibfnamefont
  {F.}~\bibnamefont {Mauri}}, \bibinfo {author} {\bibfnamefont
  {R.}~\bibnamefont {Mazzarello}}, \bibinfo {author} {\bibfnamefont
  {S.}~\bibnamefont {Paolini}}, \bibinfo {author} {\bibfnamefont
  {A.}~\bibnamefont {Pasquarello}}, \bibinfo {author} {\bibfnamefont
  {L.}~\bibnamefont {Paulatto}}, \bibinfo {author} {\bibfnamefont
  {C.}~\bibnamefont {Sbraccia}}, \bibinfo {author} {\bibfnamefont
  {S.}~\bibnamefont {Scandolo}}, \bibinfo {author} {\bibfnamefont
  {G.}~\bibnamefont {Sclauzero}}, \bibinfo {author} {\bibfnamefont
  {A.}~\bibnamefont {Seitsonen}}, \bibinfo {author} {\bibfnamefont
  {A.}~\bibnamefont {Smogunov}}, \bibinfo {author} {\bibfnamefont
  {P.}~\bibnamefont {Umari}}, \ and\ \bibinfo {author} {\bibfnamefont
  {R.}~\bibnamefont {Wentzcovitch}},\ }\href {http://www.quantum-espresso.org}
  {\bibfield  {journal} {\bibinfo  {journal} {J. Phys.: Condens. Matter}\
  }\textbf {\bibinfo {volume} {21}},\ \bibinfo {pages} {395502} (\bibinfo
  {year} {2009})}\BibitemShut {NoStop}%
\bibitem [{\citenamefont {Giannozzi}\ \emph {et~al.}(2017)\citenamefont
  {Giannozzi}, \citenamefont {Andreussi}, \citenamefont {Brumme}, \citenamefont
  {Bunau}, \citenamefont {Buongiorno~Nardelli}, \citenamefont {Calandra},
  \citenamefont {Car}, \citenamefont {Cavazzoni}, \citenamefont {Ceresoli},
  \citenamefont {Cococcioni}, \citenamefont {Colonna}, \citenamefont
  {Carnimeo}, \citenamefont {Dal~Corso}, \citenamefont {de~Gironcoli},
  \citenamefont {Delugas}, \citenamefont {Di{S}tasio~{J}r.}, \citenamefont
  {Ferretti}, \citenamefont {Floris}, \citenamefont {Fratesi}, \citenamefont
  {Fugallo}, \citenamefont {Gebauer}, \citenamefont {Gerstmann}, \citenamefont
  {Giustino}, \citenamefont {Gorni}, \citenamefont {Jia}, \citenamefont
  {Kawamura}, \citenamefont {Ko}, \citenamefont {Kokalj}, \citenamefont
  {K\"{u}\c{c}\"{u}kbenli}, \citenamefont {Lazzeri}, \citenamefont {Marsili},
  \citenamefont {Marzari}, \citenamefont {Mauri}, \citenamefont {Nguyen},
  \citenamefont {Nguyen}, \citenamefont {Otero-de-la {R}osa}, \citenamefont
  {Paulatto}, \citenamefont {Ponc\'e}, \citenamefont {Rocca}, \citenamefont
  {Sabatini}, \citenamefont {Santra}, \citenamefont {Schlipf}, \citenamefont
  {Seitsonen}, \citenamefont {Smogunov}, \citenamefont {Timrov}, \citenamefont
  {Thonhauser}, \citenamefont {Umari}, \citenamefont {Vast},\ and\
  \citenamefont {Baroni}}]{Giannozzi:2017}%
  \BibitemOpen
  \bibfield  {author} {\bibinfo {author} {\bibfnamefont {P.}~\bibnamefont
  {Giannozzi}}, \bibinfo {author} {\bibfnamefont {O.}~\bibnamefont
  {Andreussi}}, \bibinfo {author} {\bibfnamefont {T.}~\bibnamefont {Brumme}},
  \bibinfo {author} {\bibfnamefont {O.}~\bibnamefont {Bunau}}, \bibinfo
  {author} {\bibfnamefont {M.}~\bibnamefont {Buongiorno~Nardelli}}, \bibinfo
  {author} {\bibfnamefont {M.}~\bibnamefont {Calandra}}, \bibinfo {author}
  {\bibfnamefont {R.}~\bibnamefont {Car}}, \bibinfo {author} {\bibfnamefont
  {C.}~\bibnamefont {Cavazzoni}}, \bibinfo {author} {\bibfnamefont
  {D.}~\bibnamefont {Ceresoli}}, \bibinfo {author} {\bibfnamefont
  {M.}~\bibnamefont {Cococcioni}}, \bibinfo {author} {\bibfnamefont
  {N.}~\bibnamefont {Colonna}}, \bibinfo {author} {\bibfnamefont
  {I.}~\bibnamefont {Carnimeo}}, \bibinfo {author} {\bibfnamefont
  {A.}~\bibnamefont {Dal~Corso}}, \bibinfo {author} {\bibfnamefont
  {S.}~\bibnamefont {de~Gironcoli}}, \bibinfo {author} {\bibfnamefont
  {P.}~\bibnamefont {Delugas}}, \bibinfo {author} {\bibfnamefont {R.~A.}\
  \bibnamefont {Di{S}tasio~{J}r.}}, \bibinfo {author} {\bibfnamefont
  {A.}~\bibnamefont {Ferretti}}, \bibinfo {author} {\bibfnamefont
  {A.}~\bibnamefont {Floris}}, \bibinfo {author} {\bibfnamefont
  {G.}~\bibnamefont {Fratesi}}, \bibinfo {author} {\bibfnamefont
  {G.}~\bibnamefont {Fugallo}}, \bibinfo {author} {\bibfnamefont
  {R.}~\bibnamefont {Gebauer}}, \bibinfo {author} {\bibfnamefont
  {U.}~\bibnamefont {Gerstmann}}, \bibinfo {author} {\bibfnamefont
  {F.}~\bibnamefont {Giustino}}, \bibinfo {author} {\bibfnamefont
  {T.}~\bibnamefont {Gorni}}, \bibinfo {author} {\bibfnamefont
  {J.}~\bibnamefont {Jia}}, \bibinfo {author} {\bibfnamefont {M.}~\bibnamefont
  {Kawamura}}, \bibinfo {author} {\bibfnamefont {H.-Y.}\ \bibnamefont {Ko}},
  \bibinfo {author} {\bibfnamefont {A.}~\bibnamefont {Kokalj}}, \bibinfo
  {author} {\bibfnamefont {E.}~\bibnamefont {K\"{u}\c{c}\"{u}kbenli}}, \bibinfo
  {author} {\bibfnamefont {M.}~\bibnamefont {Lazzeri}}, \bibinfo {author}
  {\bibfnamefont {M.}~\bibnamefont {Marsili}}, \bibinfo {author} {\bibfnamefont
  {N.}~\bibnamefont {Marzari}}, \bibinfo {author} {\bibfnamefont
  {F.}~\bibnamefont {Mauri}}, \bibinfo {author} {\bibfnamefont {N.~L.}\
  \bibnamefont {Nguyen}}, \bibinfo {author} {\bibfnamefont {H.-V.}\
  \bibnamefont {Nguyen}}, \bibinfo {author} {\bibfnamefont {A.}~\bibnamefont
  {Otero-de-la {R}osa}}, \bibinfo {author} {\bibfnamefont {L.}~\bibnamefont
  {Paulatto}}, \bibinfo {author} {\bibfnamefont {S.}~\bibnamefont {Ponc\'e}},
  \bibinfo {author} {\bibfnamefont {D.}~\bibnamefont {Rocca}}, \bibinfo
  {author} {\bibfnamefont {R.}~\bibnamefont {Sabatini}}, \bibinfo {author}
  {\bibfnamefont {B.}~\bibnamefont {Santra}}, \bibinfo {author} {\bibfnamefont
  {M.}~\bibnamefont {Schlipf}}, \bibinfo {author} {\bibfnamefont
  {A.}~\bibnamefont {Seitsonen}}, \bibinfo {author} {\bibfnamefont
  {A.}~\bibnamefont {Smogunov}}, \bibinfo {author} {\bibfnamefont
  {I.}~\bibnamefont {Timrov}}, \bibinfo {author} {\bibfnamefont
  {T.}~\bibnamefont {Thonhauser}}, \bibinfo {author} {\bibfnamefont
  {P.}~\bibnamefont {Umari}}, \bibinfo {author} {\bibfnamefont
  {N.}~\bibnamefont {Vast}}, \ and\ \bibinfo {author} {\bibfnamefont
  {S.}~\bibnamefont {Baroni}},\ }\href@noop {} {\bibfield  {journal} {\bibinfo
  {journal} {J. Phys.: Condens. Matter}\ }\textbf {\bibinfo {volume} {29}},\
  \bibinfo {pages} {465901} (\bibinfo {year} {2017})}\BibitemShut {NoStop}%
\bibitem [{\citenamefont {{van Setten}}\ \emph {et~al.}(2018)\citenamefont
  {{van Setten}}, \citenamefont {Giantomassi}, \citenamefont {Bousquet},
  \citenamefont {Verstraete}, \citenamefont {Hamann}, \citenamefont {Gonze},\
  and\ \citenamefont {Rignanese}}]{vanSetten:2018}%
  \BibitemOpen
  \bibfield  {author} {\bibinfo {author} {\bibfnamefont {M.}~\bibnamefont {{van
  Setten}}}, \bibinfo {author} {\bibfnamefont {M.}~\bibnamefont {Giantomassi}},
  \bibinfo {author} {\bibfnamefont {E.}~\bibnamefont {Bousquet}}, \bibinfo
  {author} {\bibfnamefont {M.}~\bibnamefont {Verstraete}}, \bibinfo {author}
  {\bibfnamefont {D.}~\bibnamefont {Hamann}}, \bibinfo {author} {\bibfnamefont
  {X.}~\bibnamefont {Gonze}}, \ and\ \bibinfo {author} {\bibfnamefont {G.-M.}\
  \bibnamefont {Rignanese}},\ }\href {\doibase
  https://doi.org/10.1016/j.cpc.2018.01.012} {\bibfield  {journal} {\bibinfo
  {journal} {Comput. Phys. Commun.}\ }\textbf {\bibinfo {volume} {226}},\
  \bibinfo {pages} {39} (\bibinfo {year} {2018})}\BibitemShut {NoStop}%
\bibitem [{\citenamefont {Villar~Arribi}(2018)}]{VillarArribi:2018}%
  \BibitemOpen
  \bibfield  {author} {\bibinfo {author} {\bibfnamefont {P.}~\bibnamefont
  {Villar~Arribi}},\ }\emph {\bibinfo {title} {{Heavy fermions and Hund's
  metals in iron-based superconductors}}},\ \href
  {https://tel.archives-ouvertes.fr/tel-02151075} {\bibinfo {type} {Theses}},\
  \bibinfo  {school} {{Universit{\'e} Grenoble Alpes}} (\bibinfo {year}
  {2018})\BibitemShut {NoStop}%
\bibitem [{Note1()}]{Note1}%
  \BibitemOpen
  \bibinfo {note} {As routinely done in planewave+pseudopotential codes, atomic
  wavefunctions are obtained via a Löwdin orthogonalization of the
  pseudopotential atomic eigenstates~\cite {SanchezPortal:1995}.}\BibitemShut
  {Stop}%
\bibitem [{\citenamefont {Pizzi}\ \emph {et~al.}(2020)\citenamefont {Pizzi},
  \citenamefont {Vitale}, \citenamefont {Arita}, \citenamefont {Bl{\"u}gel},
  \citenamefont {Freimuth}, \citenamefont {G{\'e}ranton}, \citenamefont
  {Gibertini}, \citenamefont {Gresch}, \citenamefont {Johnson}, \citenamefont
  {Koretsune} \emph {et~al.}}]{Pizzi:2020}%
  \BibitemOpen
  \bibfield  {author} {\bibinfo {author} {\bibfnamefont {G.}~\bibnamefont
  {Pizzi}}, \bibinfo {author} {\bibfnamefont {V.}~\bibnamefont {Vitale}},
  \bibinfo {author} {\bibfnamefont {R.}~\bibnamefont {Arita}}, \bibinfo
  {author} {\bibfnamefont {S.}~\bibnamefont {Bl{\"u}gel}}, \bibinfo {author}
  {\bibfnamefont {F.}~\bibnamefont {Freimuth}}, \bibinfo {author}
  {\bibfnamefont {G.}~\bibnamefont {G{\'e}ranton}}, \bibinfo {author}
  {\bibfnamefont {M.}~\bibnamefont {Gibertini}}, \bibinfo {author}
  {\bibfnamefont {D.}~\bibnamefont {Gresch}}, \bibinfo {author} {\bibfnamefont
  {C.}~\bibnamefont {Johnson}}, \bibinfo {author} {\bibfnamefont
  {T.}~\bibnamefont {Koretsune}},  \emph {et~al.},\ }\href@noop {} {\bibfield
  {journal} {\bibinfo  {journal} {Journal of Physics: Condensed Matter}\
  }\textbf {\bibinfo {volume} {32}},\ \bibinfo {pages} {165902} (\bibinfo
  {year} {2020})}\BibitemShut {NoStop}%
\bibitem [{\citenamefont {Marzari}\ \emph {et~al.}(2012)\citenamefont
  {Marzari}, \citenamefont {Mostofi}, \citenamefont {Yates}, \citenamefont
  {Souza},\ and\ \citenamefont {Vanderbilt}}]{Marzari:2012}%
  \BibitemOpen
  \bibfield  {author} {\bibinfo {author} {\bibfnamefont {N.}~\bibnamefont
  {Marzari}}, \bibinfo {author} {\bibfnamefont {A.~A.}\ \bibnamefont
  {Mostofi}}, \bibinfo {author} {\bibfnamefont {J.~R.}\ \bibnamefont {Yates}},
  \bibinfo {author} {\bibfnamefont {I.}~\bibnamefont {Souza}}, \ and\ \bibinfo
  {author} {\bibfnamefont {D.}~\bibnamefont {Vanderbilt}},\ }\href {\doibase
  10.1103/RevModPhys.84.1419} {\bibfield  {journal} {\bibinfo  {journal} {Rev.
  Mod. Phys.}\ }\textbf {\bibinfo {volume} {84}},\ \bibinfo {pages} {1419}
  (\bibinfo {year} {2012})}\BibitemShut {NoStop}%
\bibitem [{\citenamefont {Souza}\ \emph {et~al.}(2001)\citenamefont {Souza},
  \citenamefont {Marzari},\ and\ \citenamefont {Vanderbilt}}]{Souza:2001}%
  \BibitemOpen
  \bibfield  {author} {\bibinfo {author} {\bibfnamefont {I.}~\bibnamefont
  {Souza}}, \bibinfo {author} {\bibfnamefont {N.}~\bibnamefont {Marzari}}, \
  and\ \bibinfo {author} {\bibfnamefont {D.}~\bibnamefont {Vanderbilt}},\
  }\href@noop {} {\bibfield  {journal} {\bibinfo  {journal} {Physical Review
  B}\ }\textbf {\bibinfo {volume} {65}},\ \bibinfo {pages} {035109} (\bibinfo
  {year} {2001})}\BibitemShut {NoStop}%
\bibitem [{\citenamefont {Miyake}\ \emph {et~al.}(2010)\citenamefont {Miyake},
  \citenamefont {Nakamura}, \citenamefont {Arita},\ and\ \citenamefont
  {Imada}}]{Miyake:2010}%
  \BibitemOpen
  \bibfield  {author} {\bibinfo {author} {\bibfnamefont {T.}~\bibnamefont
  {Miyake}}, \bibinfo {author} {\bibfnamefont {K.}~\bibnamefont {Nakamura}},
  \bibinfo {author} {\bibfnamefont {R.}~\bibnamefont {Arita}}, \ and\ \bibinfo
  {author} {\bibfnamefont {M.}~\bibnamefont {Imada}},\ }\href@noop {}
  {\bibfield  {journal} {\bibinfo  {journal} {Journal of the Physical Society
  of Japan}\ }\textbf {\bibinfo {volume} {79}},\ \bibinfo {pages} {044705}
  (\bibinfo {year} {2010})}\BibitemShut {NoStop}%
\bibitem [{\citenamefont {Graser}\ \emph {et~al.}(2010)\citenamefont {Graser},
  \citenamefont {Kemper}, \citenamefont {Maier}, \citenamefont {Cheng},
  \citenamefont {Hirschfeld},\ and\ \citenamefont {Scalapino}}]{Graser:2010}%
  \BibitemOpen
  \bibfield  {author} {\bibinfo {author} {\bibfnamefont {S.}~\bibnamefont
  {Graser}}, \bibinfo {author} {\bibfnamefont {A.~F.}\ \bibnamefont {Kemper}},
  \bibinfo {author} {\bibfnamefont {T.~A.}\ \bibnamefont {Maier}}, \bibinfo
  {author} {\bibfnamefont {H.-P.}\ \bibnamefont {Cheng}}, \bibinfo {author}
  {\bibfnamefont {P.~J.}\ \bibnamefont {Hirschfeld}}, \ and\ \bibinfo {author}
  {\bibfnamefont {D.~J.}\ \bibnamefont {Scalapino}},\ }\href {\doibase
  10.1103/PhysRevB.81.214503} {\bibfield  {journal} {\bibinfo  {journal} {Phys.
  Rev. B}\ }\textbf {\bibinfo {volume} {81}},\ \bibinfo {pages} {214503}
  (\bibinfo {year} {2010})}\BibitemShut {NoStop}%
\bibitem [{Note2()}]{Note2}%
  \BibitemOpen
  \bibinfo {note} {Here the local off-diagonal terms of the Kanamori
  Hamiltonian (spin-flip and pair-hopping) are customarily
  dropped.}\BibitemShut {Stop}%
\bibitem [{\citenamefont {de' Medici}\ \emph {et~al.}(2005)\citenamefont {de'
  Medici}, \citenamefont {Georges},\ and\ \citenamefont
  {Biermann}}]{deMedici:2005}%
  \BibitemOpen
  \bibfield  {author} {\bibinfo {author} {\bibfnamefont {L.}~\bibnamefont {de'
  Medici}}, \bibinfo {author} {\bibfnamefont {A.}~\bibnamefont {Georges}}, \
  and\ \bibinfo {author} {\bibfnamefont {S.}~\bibnamefont {Biermann}},\ }\href
  {\doibase 10.1103/PhysRevB.72.205124} {\bibfield  {journal} {\bibinfo
  {journal} {Phys. Rev. B}\ }\textbf {\bibinfo {volume} {72}},\ \bibinfo
  {pages} {205124} (\bibinfo {year} {2005})}\BibitemShut {NoStop}%
\bibitem [{\citenamefont {Hassan}\ and\ \citenamefont {de'
  Medici}(2010)}]{Hassan_CSSMF}%
  \BibitemOpen
  \bibfield  {author} {\bibinfo {author} {\bibfnamefont {S.~R.}\ \bibnamefont
  {Hassan}}\ and\ \bibinfo {author} {\bibfnamefont {L.}~\bibnamefont {de'
  Medici}},\ }\href {\doibase 10.1103/PhysRevB.81.035106} {\bibfield  {journal}
  {\bibinfo  {journal} {Phys. Rev. B}\ }\textbf {\bibinfo {volume} {81}},\
  \bibinfo {pages} {035106} (\bibinfo {year} {2010})}\BibitemShut {NoStop}%
\bibitem [{\citenamefont {de' Medici}\ and\ \citenamefont
  {Capone}(2017)}]{deMedici:2017}%
  \BibitemOpen
  \bibfield  {author} {\bibinfo {author} {\bibfnamefont {L.}~\bibnamefont {de'
  Medici}}\ and\ \bibinfo {author} {\bibfnamefont {M.}~\bibnamefont {Capone}},\
  }in\ \href@noop {} {\emph {\bibinfo {booktitle} {The Iron Pnictide
  Superconductors}}}\ (\bibinfo  {publisher} {Springer},\ \bibinfo {year}
  {2017})\ pp.\ \bibinfo {pages} {115--185}\BibitemShut {NoStop}%
\bibitem [{\citenamefont {de' Medici}(2017)}]{deMedici:2017ab}%
  \BibitemOpen
  \bibfield  {author} {\bibinfo {author} {\bibfnamefont {L.}~\bibnamefont {de'
  Medici}},\ }\href@noop {} {\bibfield  {journal} {\bibinfo  {journal}
  {Physical review letters}\ }\textbf {\bibinfo {volume} {118}},\ \bibinfo
  {pages} {167003} (\bibinfo {year} {2017})}\BibitemShut {NoStop}%
\bibitem [{\citenamefont {Chatzieleftheriou}\ \emph {et~al.}(2020)\citenamefont
  {Chatzieleftheriou}, \citenamefont {Berovi{\'c}}, \citenamefont {Arribi},
  \citenamefont {Capone},\ and\ \citenamefont {de'
  Medici}}]{Chatzieleftheriou:2020}%
  \BibitemOpen
  \bibfield  {author} {\bibinfo {author} {\bibfnamefont {M.}~\bibnamefont
  {Chatzieleftheriou}}, \bibinfo {author} {\bibfnamefont {M.}~\bibnamefont
  {Berovi{\'c}}}, \bibinfo {author} {\bibfnamefont {P.~V.}\ \bibnamefont
  {Arribi}}, \bibinfo {author} {\bibfnamefont {M.}~\bibnamefont {Capone}}, \
  and\ \bibinfo {author} {\bibfnamefont {L.}~\bibnamefont {de' Medici}},\
  }\href@noop {} {\bibfield  {journal} {\bibinfo  {journal} {Physical Review
  B}\ }\textbf {\bibinfo {volume} {102}},\ \bibinfo {pages} {205127} (\bibinfo
  {year} {2020})}\BibitemShut {NoStop}%
\bibitem [{\citenamefont {Villar~Arribi}\ and\ \citenamefont {de'
  Medici}(2018)}]{Villar-Arribi:2018}%
  \BibitemOpen
  \bibfield  {author} {\bibinfo {author} {\bibfnamefont {P.}~\bibnamefont
  {Villar~Arribi}}\ and\ \bibinfo {author} {\bibfnamefont {L.}~\bibnamefont
  {de' Medici}},\ }\href@noop {} {\bibfield  {journal} {\bibinfo  {journal}
  {Physical review letters}\ }\textbf {\bibinfo {volume} {121}},\ \bibinfo
  {pages} {197001} (\bibinfo {year} {2018})}\BibitemShut {NoStop}%
\bibitem [{\citenamefont {Sanchez-Portal}\ \emph {et~al.}(1995)\citenamefont
  {Sanchez-Portal}, \citenamefont {Artacho},\ and\ \citenamefont
  {Soler}}]{SanchezPortal:1995}%
  \BibitemOpen
  \bibfield  {author} {\bibinfo {author} {\bibfnamefont {D.}~\bibnamefont
  {Sanchez-Portal}}, \bibinfo {author} {\bibfnamefont {E.}~\bibnamefont
  {Artacho}}, \ and\ \bibinfo {author} {\bibfnamefont {J.~M.}\ \bibnamefont
  {Soler}},\ }\href@noop {} {\bibfield  {journal} {\bibinfo  {journal} {Solid
  State Communications}\ }\textbf {\bibinfo {volume} {95}},\ \bibinfo {pages}
  {685} (\bibinfo {year} {1995})}\BibitemShut {NoStop}%
\end{thebibliography}%
%%%%%%%%%%%%%%%%%%%%%%%%%%

\end{document}